\documentclass[aps,pre,singlecolumn,superscriptaddress,notitlepage]{revtex4-1}
\usepackage{amsmath,amssymb}
\usepackage{graphics,graphicx,marvosym}
\usepackage{dcolumn,bm}
\usepackage{mathbbol}
\usepackage{textpos}
\usepackage{multirow}

\newcommand{\tcmp}
{\affiliation{Condensed Matter Physics Division, 
Saha Institute of Nuclear Physics, 1/AF Bidhannagar, Kolkata 700 064 India.}}

\newcommand{\isi}
{\affiliation{Economic Research Unit, Indian Statistical Institute, 203 B. T. Road, Kolkata 700 108, India.}}

\newcommand{\isia}
{\affiliation{Sampling \& Official Statistics Unit, Indian Statistical Institute, 203 B. T. Road, Kolkata 700 108, India.}}

\graphicspath{{./figures/}}

\begin{document}
\title{Inequality in Societies, Academic Institutions and Science Journals: Gini and \textit{k}-indices}

\author{Asim Ghosh}
\email[Email: ]{asim.ghosh@saha.ac.in}
\tcmp
\author{Nachiketa Chattopadhyay}
\email[Email: ]{nachiketa@isical.ac.in}
\isia
\author{Bikas K. Chakrabarti$^{1,} $}
\email[Email: ]{bikask.chakrabarti@saha.ac.in}
\isi


\begin{abstract}
\vskip 0cm
\noindent  Social inequality is traditionally measured by the Gini-index ($g$). The $g$-index takes values from $0$ to $1$ where  $g=0$ represents complete equality and $g=1$  represents complete inequality. Most of the estimates of the income or wealth data indicate the $g$ value to be  widely dispersed across the countries of the world: \textit{g} values  typically range from $0.30$ to $0.65$  at a particular time (year). We estimated similarly  the Gini-index for the citations earned by  the yearly publications of various academic  institutions and the science journals. The ISI web of science data suggests remarkably strong inequality and universality  ($g=0.70\pm0.07$) across all the universities and institutions of the world, while for the journals we find $g=0.65\pm0.15$ for any typical year. We define a new inequality measure, namely the  $k$-index, saying that  the cumulative  income or citations of ($1-k$) fraction of people or papers  exceed those earned by  the fraction ($k$) of the people or publications respectively. We find, while the  $k$-index value for income ranges from $0.60$ to $0.75$ for income distributions across the world, it has a value around $0.75\pm0.05$ for different universities and institutions across the world and around $0.77\pm0.10$ for the science journals. Apart from above indices, we also analyze the same institution and journal citation data  by measuring Pietra index  and  median index.
\end{abstract}

\maketitle


%

\section{Introduction}
\begin{figure}[!htbp]\centering
\includegraphics[width=8cm]{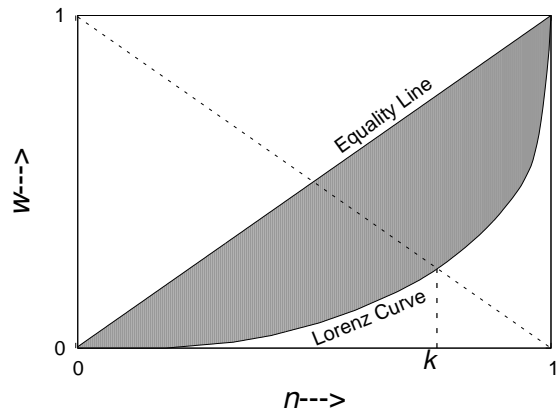}
\caption{The Lorenz curve plots the cumulative fraction $w$ of the (a) income or wealth or (b) citations when ordered from the lowest to the highest  income, wealth (for (a)) or citations (for (b))  in any particular year for any country (for (a)) or any  institution (for (b)), against the cumulative fraction (\textit{n})  of the people in the country (for (a)) or the fraction of papers published in that year (for (b)) sharing that income or wealth (for (a)) or citations (for (b)) respectively. The diagonal  starting from the origin then gives the equality line (corresponding to uniform income or wealth that year per head of the population of that country or citations per paper published that year from the institution). The (normalized) ratio of the  area of the shaded region (between the equality line and the Lorenz curve) and of the triangle formed by equality line (effectively twice the area value of the shaded region) gives the Gini-index value  ($g$). The ordinate  $k$ (on the  $n$ axis) of the intersection point of the Lorenz curve and the other diagonal (perpendicular to the equality line) gives the $k$-index. This $k$-index value gives   another measurement (similar to the $h$-index for individual scientists): $k$-value of a society  says that $1-k$ fraction of people (for (a)) or papers (for (b)) possess more  income, wealth (for (a)) or citations (for (b)) than those earned  by  the rest $k$ fraction of people (for (a)) or papers (for (b)).}
\label{sinp-2010.eps}
\end{figure}

\noindent Social inequality is  often measured by the Gini-index or Gini coefficient ($g$) \cite{gini, Coulter,Kleiber,Yitzhaki, bkc-book,cohen} obtained from the area between the diagonal (equality line) and the Lorenz curve, given by the plot of cumulative  fraction ($w$) of income or wealth (when ordered from lowest to highest income or wealth) against the corresponding  cumulative fraction ($n$) of people sharing them  in any society (at any particular time or year). In a similar way, we measure here the inequalities in the output of the various academic institutions and universities by determining  $g$-values obtained from Lorenz curves of the institutions obtained by plotting the cumulative fraction of the citations of the papers (when ordered from lowest to highest citations) published in any year from  that institution, against the corresponding cumulation fraction of papers sharing those citations (see Fig. \ref{sinp-2010.eps}).  The Gini index or $g$-value is again given by twice the (normalized) area of the region (shaded  in  Fig. \ref{sinp-2010.eps}) between the equality line or diagonal through the origin and the Lorenz curve. We introduce then a new inequality measure, the  $k$-index ($k$ for Kolkata; in view  of the extreme nature of social inequalities in Kolkata \cite{aoyama-book}) which is given  by the coordinate value  $k$ in the $n$-axis in  Fig. \ref{sinp-2010.eps} of the cutting point of the Lorenz line with the  diagonal orthogonal to the equality  line. As one can see in the case of income inequality, it says  the fraction ($1-k$) of people earn more  than that earned by  fraction  $k$ of people in the country or society. In the case of an academic institution the  $k$-value says that  the fraction $1-k$ of their papers published (in a year) from that institution have more   citations than those earned by   $k$ fraction of papers. As is obvious from Fig. \ref{sinp-2010.eps}, $g=0$ corresponds to complete equality (Lorenz curve merges with the diagonal) while $g=1$ corresponds to extreme inequity. The corresponding values of the $k$-index are $k=1/2$ for $g=0$ for complete equality and $k=1$ for $g=1$ for limiting (extreme) value of inequality. In the income or wealth inequality context, Pareto had already observed \cite{Pareto} (see also \cite{bkc-book}) that a tiny fraction (typically less than $20\%$) of (rich) people possess $80\%$ of the total wealth of the nations. The  $k$-index defined here gives a more  quantitative measure of this inequality. Also in the context of academic institutions or universities, the  $k$-index gives a (normalized)  complementary measure of the $h$-index  \cite{h-index} equivalent for the respective institution  for that year; $h$-index of a scientist gives the number  $h$ of his or her  papers,  each of which has at least $h$ citations.

Apart from $k$-index and $g$-index, we analyze these citation data for institutions and journals   by measuring   two other inequality  indices introduced recently: Pietra index or $p$-index \cite{Pietra,Eliazar1,Eliazar2} and  median index or $m$-index \cite{cohen}.  $p$-index is defined as the maximal vertical distance between the Lorenz curve and the line of perfect equality in Fig. \ref{sinp-2010.eps}. 
$m$-index is given  by  $2m^{\prime}-1$ where  $L(m^{\prime})=1/2$ ($L(x)$ denoting  the Lorentz curve). Values for  $p$-index and  $m$-index both range  from $0$ to $1$;  the value $0$  represents  complete equality while  value  $1$ represents extreme  inequality in the society. In this article, we measure inequality in income, 
citations for papers published in institutions and journals as examples for inequality in society, 
by using some newly defined quantities.

\section{Analysis and Results}

We give here a few estimated values of  $g$ and  $k$ indices of different countries in different years in table I for income inequalities across the countries of the world, with the data taken from refs. \cite{income-data1,income-data2}. In  tables II and III   we give the estimates of the values of  $g$, $k$, $p$ and $m$ indices for different institutions and universities across the world in different times (years).   These estimates are made from the corresponding Lorenz curves  drawn numerically from the respective data sets. For citations of the papers, the data are taken from ISI web of science \cite{ISI} and are counted up to the year 2013, while the publications (corresponding to the institutions of origin or to the journal where published) are for the individual years of publication (see e.g., tables II-IV).    Assuming that the Lorenz curve can be approximated by two discontinuous straight lines forming a triangle with vertex opposite to the equality line given by intersection point of the Lorenz curve and the diagonal perpendicular to the equality line, one gets $g=2k-1$.  However, this relation is very approximate and is often not satisfied for large values of $g$ and $k$.    

It is also seen that the  $k$-index value gives an upper limit for the range of fitting of power law (namely the Pareto law \cite{aoyama-book}): for $n\geq k$, $1-w\sim(1-n)^{\alpha}$ where  we find $\alpha= 0.50\pm0.10$, giving $n=1-const (1-w)^\nu$ with $\nu=2.0\pm0.5$.

\begin{table}[h]
\centering
\begin{tabular}{|l|c|c|l|c|c|}
\hline                        
Country & $g$& $k$& Country & $g$& k\\
\hline
Brazil & 0.62&  0.73 & Columbia & 0.55 & 0.70 \\

\hline 
Denmark & 0.36 &  0.63 & Finland &  0.47 &  0.67  \\

\hline 
India & 0.45  & 0.66 & Indonesia & 0.44 &  0.65 \\

\hline 
Japan & 0.31 & 0.61 & Kenya & 0.61 & 0.73 \\
\hline 
Malaysia & 0.50 & 0.68 & Netherlands & 0.44 & 0.66 \\
\hline 
NewZeland & 0.37 & 0.63 & Norway & 0.36 & 0.63\\
\hline 
Panama & 0.44 &  0.66 & SriLanka & 0.40 & 0.65\\ 
\hline 
Sweden & 0.38 &  0.64 & Tanzania & 0.53 &  0.70\\
\hline 
Tunisia &  0.50 & 0.69  & United Kingdom & 0.36 &  0.63 \\
\hline 
Uruguay & 0.49 & 0.68 &- & - &-\\
\hline 
\end{tabular}
\hspace{0.2cm}
\begin{tabular}{|l|c|c|}
\hline                        
Country & $g$& $k$\\
\hline
Australia &  0.34 & 0.62 \\
\hline 
Canada & 0.34 & 0.62\\
\hline 
Netherlands & 0.31 &   0.61  \\
\hline 
Norway &  0.31 & 0.61 \\
\hline 
Sweden &  0.29 &   0.60 \\
\hline 
Switzerland  & 0.38 &  0.63 \\
\hline 
Germany & 0.31  & 0.61 \\
\hline 
United Kingdom & 0.34  & 0.62 \\
\hline 
United States & 0.36 &  0.63  \\ 
\hline 
-&-& - \\
\hline 
\end{tabular}

\caption{$g$-index and $k$-index values for income distribution of various countries of the world during the years $1963$ to $1983$ as obtained analyzing data reported in  refs.  \cite{income-data1} (left table) and \cite{income-data2} (right table). Maximum error bar in estimated  $g$ or  $k$ values is  $\simeq0.01$.}
\end{table}

\begin{table}[!htbp]
\begin{tabular}{ |c|c|c|c|c|c|c|}
\hline
Inst./Univ. & Year & total & $g$ & $p$ & $m$ & $k$\\ 
 &  & papers/citations  &  &   & & \\ 
\hline
\multirow{4}{*}{Melbourne} & 1980 & 866/16107  & 0.67  & 0.51 & 0.79 &   0.75\\
 & 1990 &  1131/30349 &   0.68  & 0.50 & 0.82 &  0.75\\
 & 2000 & 2116/57871 & 0.65    & 0.49  &  0.78 &   0.74 \\
 & 2010 &  5255/63151  & 0.68  & 0.50 & 0.83 &  0.75  \\ \hline

\multirow{4}{*}{Tokyo} & 1980 & 2871/60682  &  0.69  & 0.52 &  0.82 &  0.76 \\
 & 1990 &   4196/108127 &  0.68  & 0.51 & 0.82 & 0.76 \\
 & 2000 & 7955/221323 & 0.70   & 0.53 & 0.84 & 0.76 \\
 & 2010 & 9154/91349   &  0.70   & 0.52 &  0.84 &  0.76 \\ \hline

\multirow{4}{*}{Harvard} & 1980 & 4897/225626  & 0.73  & 0.55 & 0.84 & 0.78\\
 & 1990 &   6036/387244 & 0.73  & 0.55 &  0.86 & 0.78 \\
 & 2000 & 9566/571666  & 0.71  & 0.54 & 0.84 & 0.77\\
 & 2010 & 15079/263600  & 0.69 & 0.52 & 0.83  & 0.76\\ \hline

\end{tabular}
\begin{tabular}{ |c|c|c|c|c|c|c|}
\hline
Inst./Univ. & Year & total & $g$ & $p$ & $m$ & $k$\\ 
 &  & papers/citations  &  &  &  & \\ 
\hline

\multirow{4}{*}{MIT} & 1980 & 2414/101929 &  0.76  & 0.59 & 0.87 & 0.79\\
 \multirow{4}{*}{} & 1990 &  2873/156707 & 0.73   & 0.56 & 0.86 & 0.78\\
 & 2000 & 3532/206165 & 0.74   & 0.56 & 0.88 & 0.78\\
 & 2010 &  5470/109995 & 0.69  & 0.51 & 0.83 &  0.76\\ \hline

\multirow{4}{*}{Cambridge} & 1980 & 1678/62981 &  0.74  & 0.56 & 0.87 & 0.78 \\
 & 1990 &  2616/111818 & 0.74   & 0.56 & 0.88 &  0.78\\
 & 2000 & 4899/196250  & 0.71  & 0.54 & 0.85 & 0.77 \\
 & 2010 &  6443/108864  &  0.70  & 0.52 & 0.85 &  0.76 \\ \hline

\multirow{4}{*}{Oxford} & 1980 & 1241/39392  & 0.70   & 0.53 & 0.83 & 0.77 \\
 & 1990 &  2147/83937 &  0.73  & 0.56 & 0.86 &  0.78 \\
 & 2000 & 4073/191096  & 0.72  & 0.54 & 0.86 &  0.77\\
 & 2010 &  6863/114657  &  0.71   & 0.53 & 0.86 &  0.76 \\ \hline

\end{tabular}

\caption{The $g$-index, $p$-index, $m$-index and $k$-index values for papers and citations (up to December 2013) of the papers published from University of Melbourne (Melbourne), University of Tokyo (Tokyo), Harvard University (Harvard), Massachusetts Institute of Technology, Cambridge University (Cambridge) and  University of Oxford (Oxford)  as obtained from ISI  web of science. The number of (total) papers and citations gives an idea about the data size involved in the analysis. The data being exact integers, there are no error in our estimated values of   the indices.}
\end{table}

\begin{table}[!htbp]
\begin{tabular}{ |c|c|c|c|c|c|c|}
\hline
Inst./Univ. & Year & total & $g$ & $p$ & $m$ & $k$ \\ 
 &  & papers/citations  &  &  & &  \\ 
\hline
\multirow{4}{*}{SINP} & 1980 & 32/170 & 0.72  &  0.49 & 0.87&  0.74\\
\multirow{4}{*}{} & 1990 & 91/666  & 0.66   &   0.47 & 0.82 & 0.73\\
 & 2000 & 148/2225 & 0.77  & 0.57 & 0.93  & 0.79\\
 & 2010 &   238/1896  & 0.71   & 0.52 & 0.89 & 0.76\\ \hline

\multirow{4}{*}{IISC} & 1980 & 450/4728  & 0.73   & 0.56 & 0.84 & 0.78\\
\multirow{4}{*}{} & 1990 & 573/8410  & 0.70   & 0.53 &  0.83 &  0.76\\
 & 2000 & 874/19167 & 0.67   & 0.50 & 0.81 & 0.75\\
 & 2010 & 1624/11497& 0.62  & 0.45 & 0.76 & 0.73 \\ \hline

\multirow{4}{*}{TIFR} & 1980 & 167/2024 &  0.70    & 0.52 &  0.83 & 0.76 \\
 \multirow{4}{*}{}& 1990 & 303/4961 &  0.73  & 0.54 &  0.89  & 0.77\\
 & 2000 & 439/11275 &  0.74   & 0.55 & 0.90  & 0.77 \\
 & 2010 & 573/9988 &  0.78    & 0.59 & 0.95 & 0.79 \\ \hline

\end{tabular}
\begin{tabular}{ |c|c|c|c|c|c|c|c|}
\hline
Inst./Univ. & Year & total & $g$ &  $p$ & $m$ & $k$\\ 
 &  & papers/citations  &  &  & &  \\ 
\hline
\multirow{4}{*}{Calcutta } & 1980 & 162/749 &  0.74   & 0.56 & 0.86 &  0.78\\
 & 1990 &  217/1511  & 0.64  & 0.48  & 0.74 &  0.74\\
 & 2000 & 173/2073 & 0.68  & 0.50 & 0.80 &  0.74\\
 & 2010 & 432/2470  & 0.61  & 0.45 & 0.73 & 0.73\\ \hline

\multirow{4}{*}{Delhi } & 1980 &  426/2614  & 0.67   & 0.49  & 0.80 & 0.75\\
 & 1990 &  247/2252  &  0.68  & 0.52 & 0.81 &  0.76\\
 & 2000 &   301/3791 & 0.68  &  0.51 & 0.81 &  0.76 \\
 & 2010 & 914/6896 & 0.66   & 0.49 & 0.80 & 0.74\\ \hline

\multirow{4}{*}{Madras} & 1980 & 193/1317 &  0.69   & 0.53 &  0.78 & 0.76\\
 & 1990 & 158/1044 & 0.68   & 0.52 & 0.80 &  0.76 \\
 & 2000 &  188/2177 & 0.64   & 0.47 & 0.78 &  0.73 \\
 & 2010 & 348/2268  & 0.78  & 0.58 & 0.92 & 0.79 \\ \hline
\end{tabular}

\caption{The $g$-index, $p$-index, $m$-index and $k$ index  values for  Indian institutions Saha Institute of Nuclear Physics (SINP), Indian Institute of Science (IISC), Tata Institute of Fundamental Research (TIFR), Calcutta University (Calcutta), Delhi University (Delhi) and Madras University (Madras). As in table II, all the data are obtained from ISI web of science.}
\end{table}

\begin{table}[!htbp]
\begin{tabular}{ |c|c|c|c|c|c|c| }
\hline

Journals  & Year & total & $g$ & $p$ & $m$ & $k$\\ 
 &  & papers/citations  &  &  & & \\ 
\hline
\multirow{4}{*}{Nature} & 1980 & 2904/178927  & 0.80 & 0.63 & 0.89 &   0.81  \\
 & 1990 & 3676/307545  &  0.86  & 0.72 & 0.92 &   0.85\\
 & 2000 & 3021/393521  & 0.81   & 0.65 & 0.89 &  0.82\\
 & 2010 &  2577/100808  & 0.79  & 0.63 & 0.86 & 0.81\\ \hline

\multirow{4}{*}{Science} & 1980 & 1722/111737 & 0.77   & 0.60 &  0.87 & 0.80\\
 & 1990 & 2449/228121    & 0.84   & 0.70 & 0.90  & 0.84\\
 & 2000 & 2590/301093    & 0.81    & 0.66 & 0.88  & 0.82\\
 & 2010 &  2439/85879    & 0.76   & 0.60 & 0.84  & 0.79\\ \hline

\multirow{4}{*}{PNAS} & 1980 & -  &  - &  - & - & - \\
\multirow{4}{*}{(USA)}  & 1990 & 2133/282930 & 0.54   & 0.39 & 0.70 &  0.70\\
 & 2000 &  2698/315684  & 0.49  & 0.35 & 0.63 & 0.68\\
 & 2010 & 4218/116037  &   0.46  & 0.33 & 0.69 &  0.66 \\ \hline

\multirow{4}{*}{Cell} & 1980 & 394/72676 & 0.54   & 0.39 & 0.70 &   0.70\\
  & 1990 &  516/169868  & 0.50   & 0.36 & 0.65 &   0.68\\
 & 2000 &  351/110602  & 0.56  & 0.41 & 0.74  & 0.70\\
 & 2010 &  573/32485 &   0.68   & 0.51 & 0.79 &  0.75 \\ \hline

\hline

\multirow{4}{*}{PRL} & 1980 & 1196/87773 & 0.66  & 0.48 & 0.84 & 0.74\\
\multirow{4}{*}{}  & 1990 &  1904/156722  & 0.63  &  0.47 & 0.78 &  0.74\\
 & 2000 & 3124/225591  & 0.59  & 0.43 & 0.74  &  0.72\\
 & 2010 & 3350/73917  & 0.51    & 0.37 & 0.66 &  0.68 \\ \hline

\end{tabular}
\begin{tabular}{ |c|c|c|c|c|c|c| }
\hline

Journals  & Year & total & $g$ & $p$ & $m$ & $k$\\ 
 &  & papers/citations  &  & & & \\ 
\hline

\multirow{4}{*}{PRA} & 1980 & 639/24802 & 0.61  & 0.45 &  0.77 & 0.73 \\
 & 1990 & 1922/54511   & 0.61   & 0.45 & 0.76 &  0.72\\
 & 2000 &  1410/38948  &  0.60  & 0.44 & 0.77 &  0.72 \\
 & 2010 &  2934/26314  &  0.53  & 0.38 & 0.68  &  0.69 \\ \hline

\multirow{4}{*}{PRB} & 1980 & 1413/62741   &  0.65   & 0.49 & 0.81 & 0.74 \\
 & 1990 & 3488/153521  & 0.65   & 0.48 & 0.81 & 0.74 \\
 & 2000 &  4814/155172  &  0.59   & 0.44 & 0.75 & 0.72 \\
 & 2010 &  6207/70612  &  0.53   & 0.38 & 0.68 &  0.69\\ \hline

\multirow{4}{*}{PRC} & 1980 & 630/19373  & 0.66   & 0.49 & 0.82 &  0.75\\
 & 1990 &  728/15312   & 0.63   &  0.46 & 0.79 &  0.73 \\
 & 2000 & 856/19143   &  0.57   & 0.42 & 0.72 &  0.71 \\
 & 2010 & 1061/11764  &  0.56  & 0.40 & 0.70 &  0.70\\ \hline

\multirow{4}{*}{PRD} & 1980 &  800/36263 & 0.76   & 0.59 & 0.90 &  0.80\\
 & 1990 & 1049/ 33257   &  0.68  & 0.52 & 0.82 &  0.76\\
 & 2000 &  2061/66408  & 0.61   & 0.45 & 0.76 &  0.73 \\
 & 2010 & 3012/40167  &  0.54   &  0.39 & 0.68 & 0.69\\ \hline

\multirow{4}{*}{PRE} & 1980 &  -  &  -  &  -& -&- \\
 & 1990 &  -   &  - &  - & -&-  \\
 & 2000 &  2078/ 51860 &  0.58    & 0.42 & 0.73 &  0.71\\
 & 2010 & 2381/16605 & 0.50  & 0.36 & 0.63  & 0.68\\ \hline

\end{tabular}

\caption{The $g$-index, $p$-index, $m$-index and $k$ index  values for papers and citations (up to December 2013) of the papers published from different Journals,  as obtained from ISI  web of science.}
\end{table}

\begin{figure}[!htbp]\centering
\includegraphics[width=8cm]{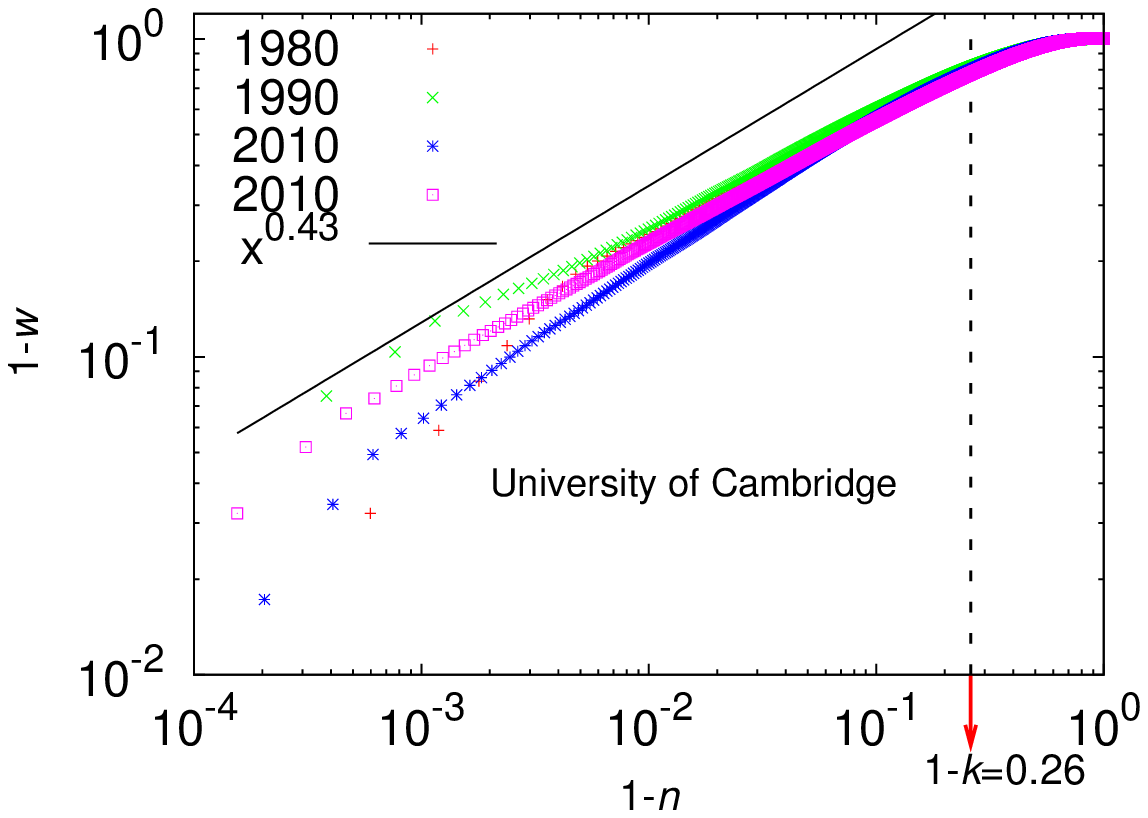}
\includegraphics[width=8cm]{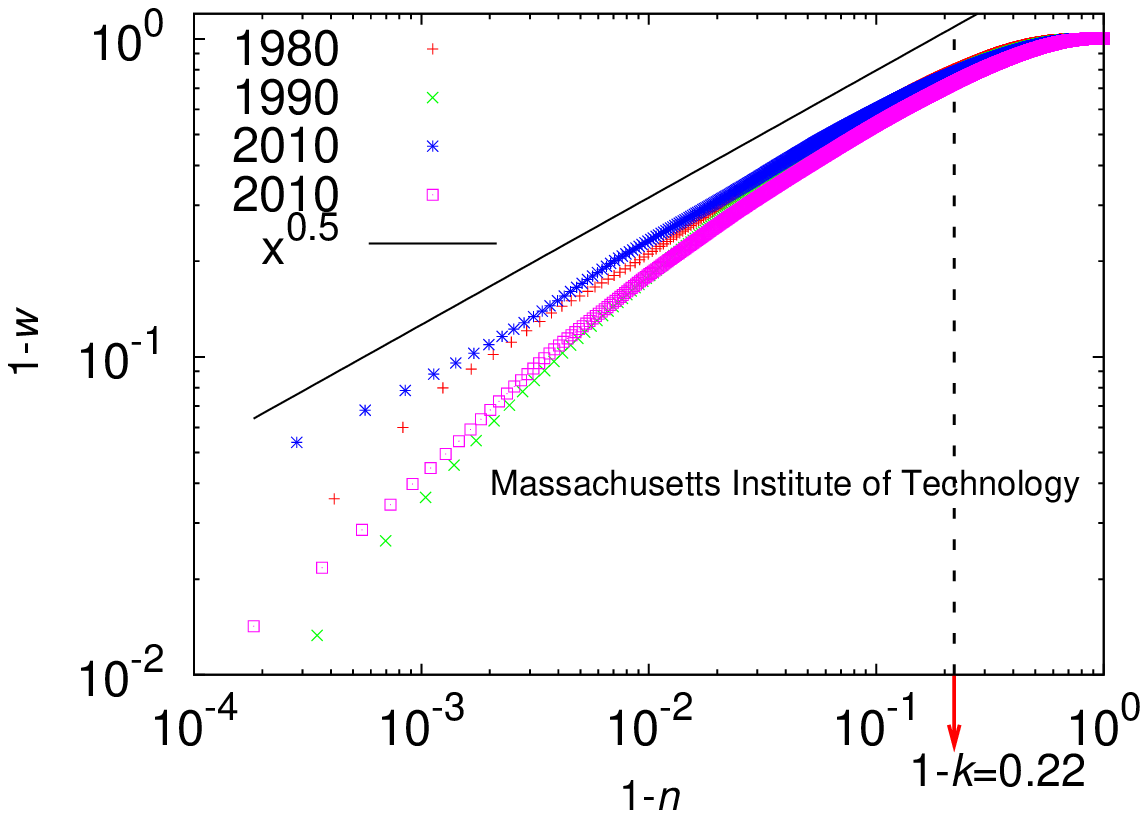}
\caption{Plot of $1-w$ against $1-n$ for citation distributions for a few institutions, showing that their variation with the corresponding  publication numbers follow a Pareto type power law behavior  beyond the $k$-index value of $n$: $1-w\sim(1-n)^\alpha$ for $n\geq k$, with $\alpha=0.50\pm0.10$ .}
\label{power-law}
\end{figure}

\begin{figure}[!htbp]\centering
\includegraphics[width=8cm]{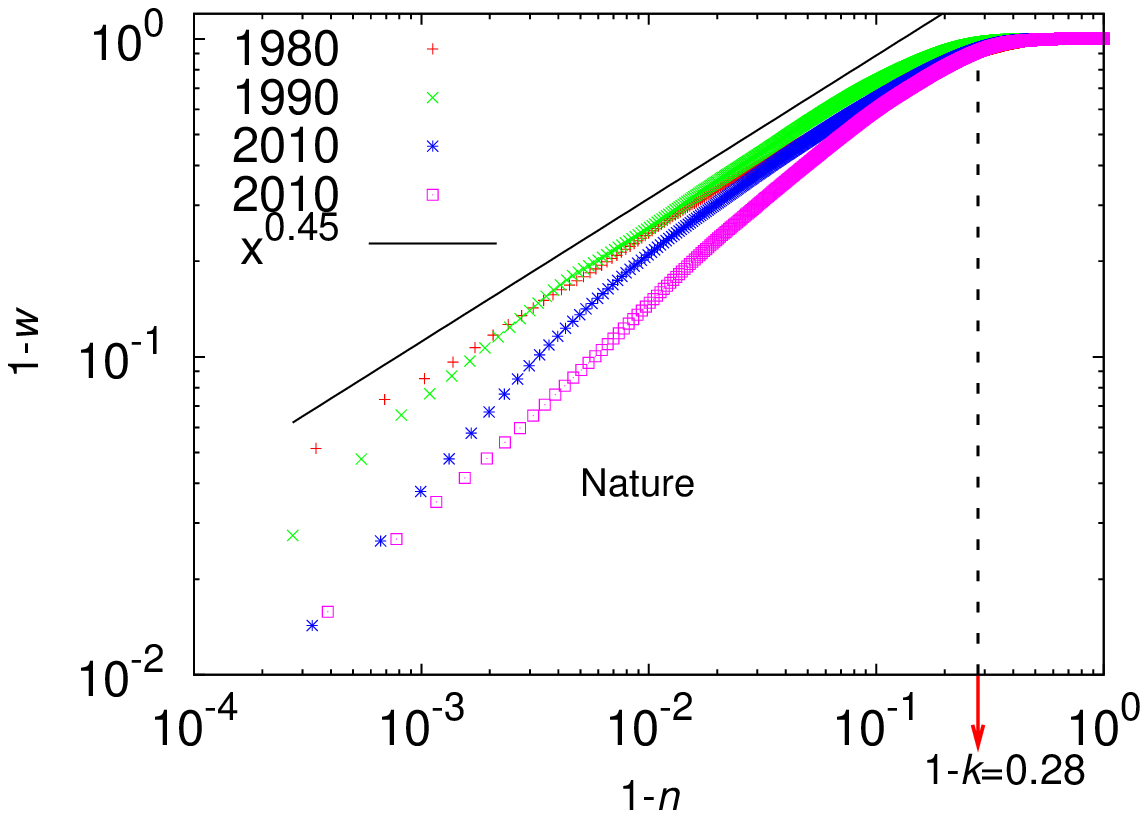}
\includegraphics[width=8cm]{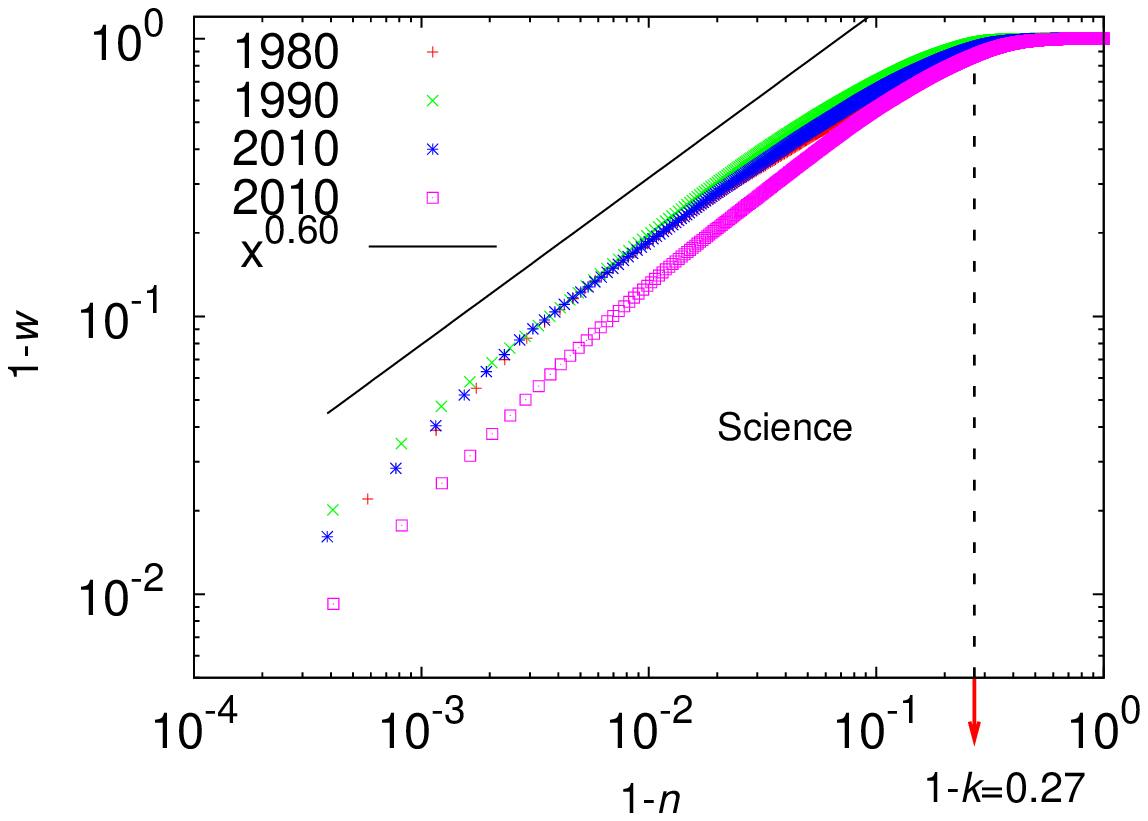}
\caption{Plot of $1-w$ against $1-n$ for citation distributions for a few science journals, showing that their variation with the corresponding  publication numbers follow a Pareto type power law behavior  beyond the $k$-index value of $n$: $1-w\sim(1-n)^\alpha$ for $n\geq k$, with $\alpha=0.50\pm0.10$ .}
\label{power-law2}
\end{figure}

\section{Summary and Conclusion}
Social inequality is traditionally measured by the Gini-index ($g$). Recently, a few more indices   measuring  social   inequality have been introduced:  $p$-index \cite{Pietra} and $m$-index \cite{cohen} as discussed earlier. It may be noted that $g$, $p$ and $m$  take values within the range $0$ (representing complete equality) and  $1$ (representing complete inequality).  We introduced the $k$-index here, signifying that $1-k$ fraction of people or papers earn more wealth or citations than the rest $k$ fraction of people or papers. As such, the lowest value of $k$-index is $0.5$ (complete equality) and the highest value is $1$ (corresponding to complete inequality). $k$-index can be  rescaled to unit interval via the  transformation $k\Rightarrow2k-1$, where   $2k-1$ gives  the vertical distance at the point $k$, between the  perfect equality line  and the  Lorentz curve.

 Most of the estimates of the income or wealth data indicate the $g$ value to be  the widely dispersed across the countries of the world: \textit{g} values  typically range from $0.30$ to $0.75$   at any particular time or year. We estimated similarly  the Gini-index for the citations earned by  the yearly publications of various academic  institutions. The ISI web of science data suggests remarkably strong inequality and universality  ($g\simeq0.70\pm0.05$) across all the universities and institutions of the world (see also \cite{Fortunato}). We   also find  that   most of the  $p$-index values for universities and institutions range from $0.40$ to $0.60$   and similarly  for $m$-index it is ranges from $0.80$ to $0.90$. We define here  a new inequality measure, namely the  Kolkata-index or  $k$-index and  find that  while the  $k$-index value for income distributions  ranges from $0.60$ to $0.75$  across the world, it has a value around $0.75\pm0.05$ for different universities and institutions across the world.  As such, $k$-index is the social equivalent to the $h$-index for an individual researcher in science. Also we find   that the value for $k$-index gives an estimate of  the crossover point beyond which the growth of income (or citations) with the fraction of population (or publications) enters a power law (Pareto) region.

\section*{Acknowledgement}
We would like to thank  Anindya S. Chakrabarti and Arnab Chatterjee for some useful comments and suggestions.


\end{document}